\documentclass[useAMS,usenatbib]{mn2e}
\usepackage{graphicx}

\title[The distance to the Sgr dwarf spheroidal galaxy]
{The distance to the Sgr dwarf spheroidal galaxy from the 
Red Giant Branch Tip}
\author[L. Monaco et al.]{L. Monaco$^{1,2,3}$, M. Bellazzini$^{2}$, 
F.R. Ferraro$^{3}$, E. Pancino$^{2}$
\thanks{E-mail: lmonaco@ts.astro.it; michele.bellazzini@bo.astro.it, francesco.ferraro3@unibo.it, elena.pancino@bo.astro.it}\\
$^{1}$INAF - Osservatorio Astronomico di Trieste, Via Tiepolo 11, 34131, Trieste, 
Italy\\
$^{2}$INAF - Osservatorio Astronomico di Bologna, via Ranzani 1, 40127, Bologna,
Italy\\
$^{3}$Universit\`a di Bologna - Dipartimento di Astronomia, via Ranzani 1, 40127, Bologna,
Italy}

\begin{document}

\date{\today}

\pagerange{\pageref{firstpage}--\pageref{lastpage}} \pubyear{2003}

\maketitle

\label{firstpage}

\begin{abstract} 

We derived the distance to the central region of the Sagittarius dwarf
spheroidal galaxy from the Red Giant Branch
Tip. The obtained distance modulus is $(m-M)_0=17.10\pm0.15$, corresponding
to a heliocentric distance  $D=26.30\pm1.8$ Kpc. This 
estimate is in good agreement with the distance obtained from RR Lyrae stars of
the globular cluster M~54, located in the core of the Sgr galaxy, once the most
accurate estimate  of the cluster metallicity and the most recent calibration
of the $M_V(RRLy)~vs.[Fe/H]$ relation are adopted.   \end{abstract}


\section{Introduction}

The Sagittarius dwarf spheroidal galaxy (Sgr dSph) is a nearby satellite of the
Milky Way \citep{s1,s2} that is currently disrupting under the strain of the
Galactic tidal field. The relic of the process of  disruption is observed as a
huge coherent structure of stars escaped by the main body of the galaxy that
remains approximately aligned along the original orbital path 
\cite[the Sgr Stream, see][and references therein]{orb-sds,orb-2mass,sdss,mb2,steve}. 
This occurrence provides an unprecedented occasion to study in detail
the mechanism of merging of galactic sub-units as well as to constrain the mass 
and the {\em shape} of the dark matter halo of the Milky Way \citep{orb-sds}, 
and it has triggered a burst of theoretical efforts 
to accurately model the dynamical evolution of the Sgr system 
\cite[see, for example][]{velas,orb,katy,gomez,binney}. 

A fundamental ingredient of any realistic model of the dynamical history of Sgr
is its distance, a key parameter that is not so well constrained, at present.
The available estimates range from $(m-M)_0=16.90 \pm 0.15$ \citep{ala01}
to $(m-M)_0=17.25^{+0.10}_{-0.20}$ \citep{sdgs1}. Other estimates
have been provided, among the others, by 
\citet{musk,s2,marc97,ls00}. While all these distance moduli are formally 
compatible within the errors, they allow distances from $D=22.4$ Kpc to
$D=29.5$~Kpc. It has also to be
considered that, among the above analysis, the only one that was properly
designed to obtain
an accurate distance estimate is that by \citet[][hereafter LS00]{ls00}. 
These authors studied the RR Lyrae population of the globular cluster M~54, that
is located in the core of the Sgr dSph. From a robust estimate of the average
apparent magnitude of the {\em ab} type RR Lyrae of M~54 ($<V_{RR}>=18.17\pm0.01$)
they obtained $(m-M)_0=17.19\pm 0.12$.

In this contribution we provide a new estimate of the distance to the Sgr
galaxy using the Tip of the Red Giant Branch (TRGB) \citep[see][for details and
references about the method]{lfm93,smf96,mf98,scw}. We measured the apparent
magnitude of the TRGB in the I passband and we derived the respective distance
modulus adopting our  recent calibration of the method \citep{tip1,tip2}, whose
zero-point is fully independent from the RR-Lyrae distance scale  \cite[that is
still affected by sizeable uncertainties, see][]{carla,walk}.  

\section{Detection of the TRGB}\label{detection} 

\begin{figure}
\includegraphics[width=84mm]{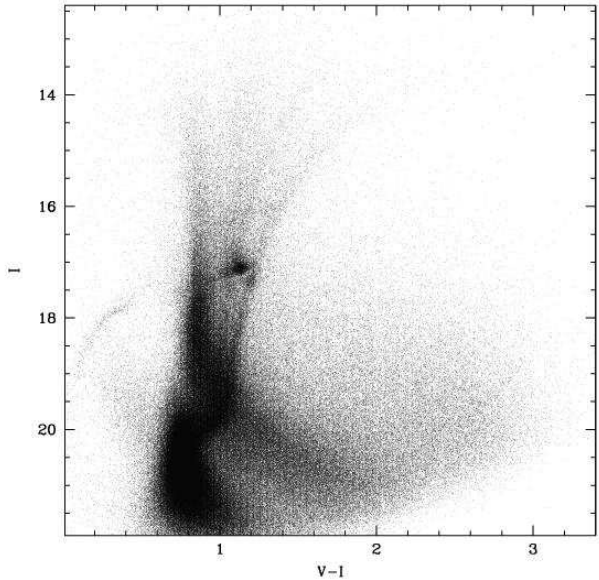} 
\caption{I {\it vs} V-I CMD for a $1\degr \times 1\degr$ field centered on
M~54.}
\label{1pre}
\end{figure}

In order to measure the magnitude of the TRGB, we used the  huge catalogue of
$\sim$490,000  stars obtained in a $1\degr \times 1\degr$ field centered on
M~54 presented by \citet{bump} and plotted in Fig.~\ref{1pre}.

\begin{figure}
\includegraphics[width=84mm]{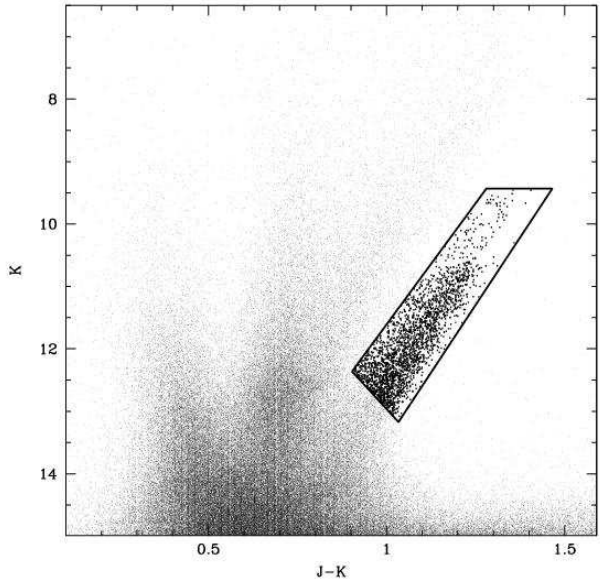} 
\caption{K {\it vs} J-K CMD for a $4\degr \times 4\degr$ 
field centered on M~54. 
The box encloses stars belonging to the Sgr's RGB.}
\label{1}
\end{figure}

\begin{figure}
\includegraphics[width=84mm]{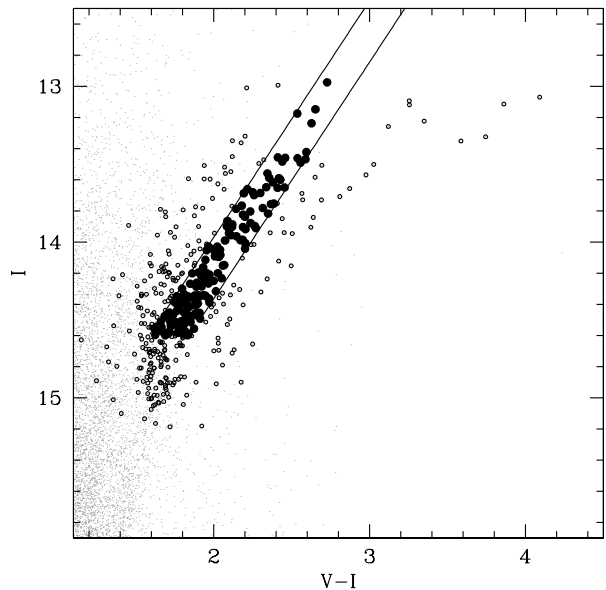} 
\caption{Zoomed I {\it vs} V-I  CMD in the RGB region (light gray). 
Stars selected in Fig~\ref{1} and also measured in the optical bands 
are marked. The thick parallel line shows how the RGB stars has been selected
for the detection of the TRGB. Stars belonging to the final selection are 
plotted as large filled circles.}
\label{1post}
\end{figure}

The Red Giant Branch (RGB) stars in Sgr can be more easily separated from
the contaminating bulge and disc sequences \citep[see, for example,][]{co01} 
in the infrared Color Magnitude Diagram (CMD).
Therefore, we
used the infrared database from the 2MASS survey to select our RGB sample. We
extracted Near Infrared (NIR) photometry of a $4\degr \times 4\degr$ field
centered on M~54 from the Point Source Catalogue of the All Sky Data Release of
2MASS \citep{cutri}. We selected only high quality sources, avoiding any
possible blended stars and/or contaminants, as done in \citet{mb2}. The
resulting sample of {\it bona fide}  Sgr stars is shown in Fig.~\ref{1},
enclosed in a box.
In Fig.~\ref{1post} we plotted the  I {\it vs} V-I CMD zoomed in the RGB
region. Stars selected in Fig.~\ref{1}   and also measured in the optical
bands  are marked. 

It is well known that the largely dominant stellar component of the Sgr dSph
galaxy is a population of quite high mean metallicity ($-0.6 \la$
[M/H]\footnote{The ``global metallicity'' is defined as:
[M/H]=[Fe/H]+log(0.638$\times$10$^{[\alpha/Fe]}$+0.362), see \citet{scs93}} 
$\la -0.4$) and of age  $\sim 6$ Gyr \citep{sdgs2,ls00,co01,bump}. 
Nevertheless the {\it bona-fide} Sgr member in Fig~\ref{1post} show a large
color spread in the CMD, witnessing the presence of stars of different ages and
metallicity \citep{musk,sdgs2,ls00,bhbletter,boni04}.  Since the absolute
magnitude of the Tip is (weakly) dependent on metallicity, it would be safe  to
exclude stars  not related to the Sgr dominant population from the sample used
for the detection  of the TRGB.

Therefore, we compared the optical CMD of Sgr with  the RGB ridge lines of
template globular clusters    \cite[see][]{bump} and we finally adopted the
selection shown in Fig.~\ref{1post}. According to our experiments with ridge
lines, the adopted selection (the stars included between the two parallel thick
lines in the plot) likely excludes stars with $[M/H]\la -1.0$ and $[M/H]\ga
0.0$. In this way the bulk of the dominant population \cite[whose age and
metallicity range is rather limited, see][]{bump} is picked up by our
selection.

\begin{figure}
\includegraphics[width=84mm]{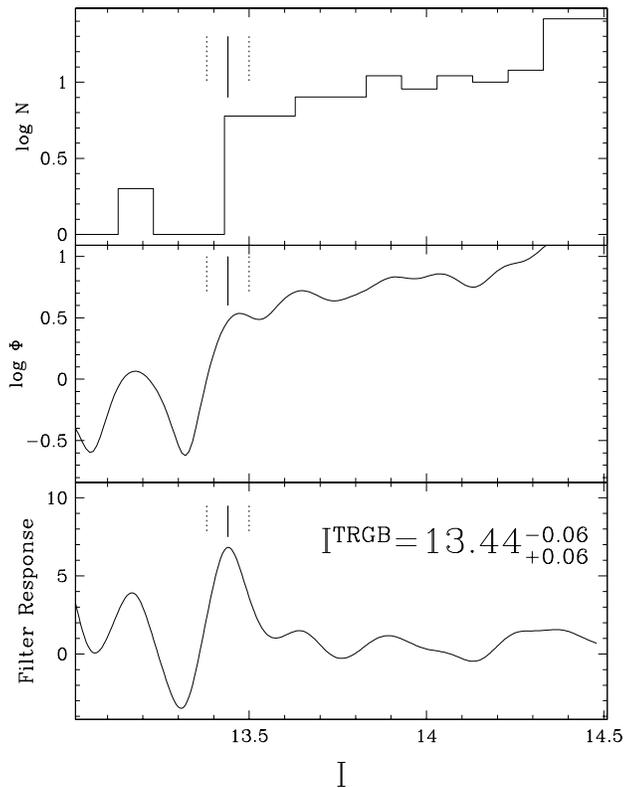} 

\caption{Upper panel: Logarithmic LFs of the upper
RGB for the stars selected in Fig.~\ref{1post}. The position of the TRGB is
marked by a thick vertical line, the thin lines enclose the error bars.  Middle
panel:  Logarithmic LF of the upper RGB as a generalized histogram. 
Lower panel: Sobel filter response to the LFs as a
function of the I magnitude. The obvious peaks indicate the magnitude of the
TRGB.}
\label{2}
\end{figure}

The observable signature of the TRGB is the sharp cut-off at the bright end of
the RGB Luminosity Function (LF). In well populated LFs the cut-off is easily
detected applying an edge-detector filter (i.e. the Sobel Filter) to the LF
\cite[see][and references therein]{lfm93,mf95,mf98,tip1}. The detection of the
TRGB in Sgr in the I band is shown in Fig.~\ref{2}. The LF cut-off is evident
in the LF shown in the upper and middle panels \citep[presented as a
generalized histogram in the latter, see][]{laird} and it is recognized as
an obvious maximum in the filter response (lower panel).  The reported
uncertainties are the Half Widths at Half Maximum (HWHM) of the peaks of the
filter response.  The final estimate is $I^{TRGB}=13.44\pm0.06$. We note that
there are  $N_{\star}=121$ stars within one magnitude from the detected TRGB,
hence the sampling criterion for a safe estimate of the TRGB distance is
fulfilled in the present case  \cite[i.e. $N_{\star}>100$, see][]{mf95,b02}.

\section{The distance to the Sgr dSph} 

The distance to the Sgr dSph is derived from the TRGB by using the calibrating
relations from \citet[][hereafter B04]{tip2}.  In that paper we provided
empirical relations for the absolute magnitude of the TRGB in I,J,H, and K as a
function of the global metallicity of the considered stellar system. The global
metallicity ([M/H]) is a parameter that takes into account not only the
abundance of iron, but also of the $\alpha$-elements (O, Mg, Si, etc.), making
easier and more self-consistent the comparison between systems having different
abundance patterns and with theoretical models  \cite[see][and B04 for
definitions and details]{scs93,f99}. The zero-point of the adopted calibration
is based on the distance estimate to the globular cluster $\omega$ Centauri
obtained by \citet{ogle} from the double-line detached eclipsing binary OGLE-17
\cite[see][B04]{tip1}, hence it is fully independent from the classical
distance scale based on RR Lyrae and/or Horizontal Branch stars in general 
\cite[see][for applications and discussions]{tosi2,b02,apella,alves,
alves2,walk}. 

Assuming $E(B-V)=0.14$ \citep[see][]{ls00} and a mean metallicity of [M/H]=-0.5
\citep[see][]{bump} for the Sgr dominant population, the measured
$I^{TRGB}=13.44$ corresponds to:  $(m-M)_0$=17.10.

The final distance estimate is affected by the uncertainties in:
\begin{itemize}
\item 	the {\bf calibrating relations:} the uncertainty of the zero point of the 
	calibrations as estimated by
      	B04 is $\sigma=0.12$ mag for the I passband. 
\item	the {\bf metallicity} of the Sgr population: according to the relations in B04, 
	a variation of $\pm$0.1~dex in 
	the mean metal content produce a change in the absolute I magnitude 
	of the TRGB of about 0.045~mag. 
\item	the estimate of the {\bf apparent TRGB magnitude} which has an 
	uncertainty of $\pm$0.06~mag.  
\item	{\bf the reddening:} we adopt the
	direct reddening estimate obtained by LS00 from the color at the minimum of the
	light curve of {\em ab} RR Lyrae: $E(B-V)=0.14\pm 0.03$.  This is in good
	agreement with the results from the reddening maps of \citet{iras}, once the
	corrections provided by \citet{dutra} are taken into account. Inspection of
	the reddening maps provided also strong indications that there is no serious
	variability of extinction in the considered field. The standard	
	deviation of the reddening value, interpolated from the map at the position of
	each star in the 2MASS field, is just $\sigma_{E(B-V)}=0.03$, e.g. of the order
	of the uncertainty in the reddening estimate. Concerning the reddening laws, we
	adopt $A_I=1.76E(B-V)$ (Cousins' I) from \citet{dean}.
\end{itemize}
All these uncertainty factors are independent each other. Therefore they can be quadratically summed 
to produce an average error of 0.15~mag, of the order of that abtained in other previous estimates 
of the distance to Sgr. 

In summary, the derived
distance modulus is $(m-M)_0=17.10\pm0.15$, corresponding to a heliocentric
distance  $D=26.30\pm1.8$ Kpc.

In comparison to the infrared magnitudes, the I magnitude of the  TRGB has a
weaker dependence on metallicity. Moreover, the current calibrating relations
suffer by a higher uncertainty at the infrared wavelenghts. Therefore, it is
certainly safer to use the I magnitude of the TRGB for the  derivation of the
distance. Nevertheless a comparison with the infrared database can provide a
useful consistency check for the derived distance modulus.

\begin{figure}
\includegraphics[width=84mm]{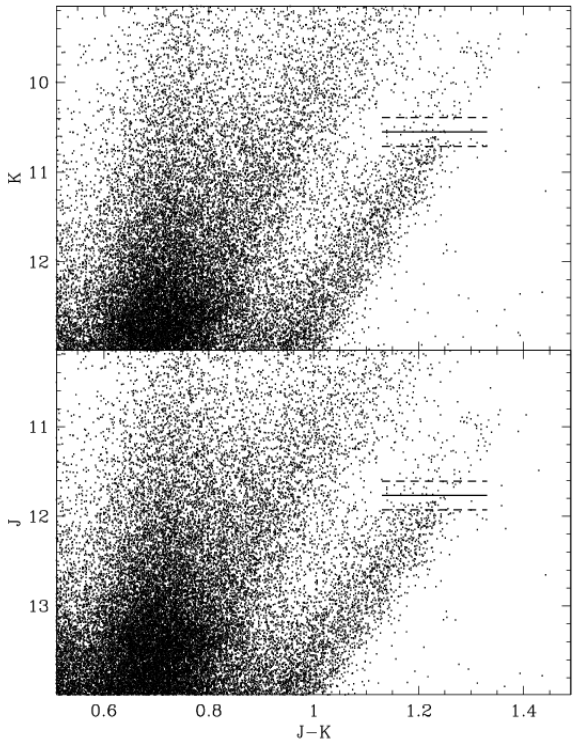} 
\caption{K {\it vs} J-K (upper panel) and J {\it vs} J-K (lower panel) CMDs. The
continuous horizontal lines represent the position of the TRGB derived from our
(m-M)$_0$ and the calibrating relations by B04 assuming a mean metallicity of
[M/H]=-0.6.}
\label{4}
\end{figure}

The 2MASS infrared photometry already introduced in  section \ref{detection} is
plotted in figure \ref{4} in the K {\it vs} J-K (upper panel) and J {\it vs}
J-K (lower panel) planes. We also plotted on each plane the {\it predicted}
magnitude of the TRGB (middle continuous line) with the appropriate uncertainty
(top and bottom lines). The predicted magnidude of the TRGB in the J and K
passbands have been obtained from the derived distance modulus using   the
calibration provided in B04. The calibrations by B04 are in the same
photometric system adopted by  \citet[][hereafter F00]{f00} and we used the
empirical relations provided by \citet{valenti} to convert the F00 into the
2MASS system:  $J_{2MASS}=J_{F00}-0.06$ and  $K_{2MASS}=K_{F00}-0.05$.  We also
adopted E(B-V)=0.14 (LS00) and the reddening laws from \citet{rieke}:
$A_J=0.871E(B-V)$ and  $A_K=0.346E(B-V)$. 

We find that, by slightly adjusting the global metallicity to [M/H]=-0.6
\citep[which is still consistent with the current estimates,
see][]{bump,ls00}, a fairly good agreement between the predicted and
observed position of the TRGB is reached in the infrared passbands 
(see Fig.~\ref{4}) which proves the internal consistency in  the calibrating relations
of B04.

\section{Discussion}

We have provided a new estimate of the distance modulus to the Sgr dSph galaxy
using the TRGB in the I passband as a standard candle.
The obtained  distance modulus is
$(m-M)_0=17.10\pm0.15$ which corresponds to a heliocentric distance 
$D=26.30\pm1.8$ Kpc.

Due to the large uncertainties present in the various studies, this estimate is
formally compatible with all existing measures.  However, it is worth noting
that  most of existing distance estimates are often based on crude assumptions
and many of them are likely affected by various kinds of systematics. In many
cases the accuracy of these estimates was also hampered by the limited
knowledge of the physical parameters (e.g. metallicity, age, etc.) of the main
population of Sgr. 

In the earliest studies the distance was derived by comparing the mean 
magnitude of the Red Clump of Sgr with that of template globular cluster or
nearby galaxiess  \citep[][]{sdgs1,marc97,igi}, e.g. neglecting the 
significant dependence  of the RC on the age of the considered population 
\citep[see][and references therein]{rc1}. Other authors \citep{s2,sl95}
extrapolated the mean magnitude of the RR Lyrae (M$_V$(RR)) from the RC
magnitude, adding a further source of uncertainty, and derived the distance
modulus from a   M$_V$(RR) {\it vs} [Fe/H] relation, carrying its own
uncertainty.  The actual metallicity of the RR Lyrae stars of Sgr is poorly
known  \citep[but see][]{cser01}. This fact limited the accuracy of distance
estimates directly derived from the properties of the RR Lyrae \citep{mateo2}. 
For example, \citet{musk,ala01,cser} assumed a constant M$_V$(RR),  completely
neglecting the variation of the RR Ly luminosity with metallicity. \citet[][]{m55}
identified Sgr main sequence stars in the background of the  globular cluster
M~55. They derived a range of distance moduli by fitting the sequence with
theoretical isochrones under different assumptions on the age and  metallicity
of Sgr. However too few stars were present to obtain an accurate fit and  the
age and metallicity adopted are quite different from what currently  assumed
for  the dominant population of Sgr \citep[see][and references therein]{bump}.
The result presented here is largely free from the systematics affecting the
above described analysis.

The most accurate determination of the distance to Sgr available up to now is 
the one provided by LS00, based on the RR Lyrae of the globular cluster M~54.
M~54 resides in the densest
clump of Sgr stars \citep{sl95,nucleo} and it is part of the Sgr galaxy.  
From the mean magnitude of the {\em ab} RR Lyrae of M~54, LS00 derived
$(m-M)_0=17.19\pm0.12$ as distance to M~54 and, hence, to Sgr. In particular
they measured a $<V^{RR}_0>=17.74\pm0.03$ and assumed $[Fe/H]=-1.79$ for M~54
from their previous photometric study \citep{sl95}. Then by considering the
$M_V(RR)~vs.~[Fe/H]$ relation of \citet{chab} they derived the distance
modulus. The error in the distance modulus, 0.12, reflects the uncertainty in
the zero point of the $M_V(RR)~vs.~[Fe/H]$ relation.

However, the high-resolution spectroscopic study of M~54 stars by \citet{bwg}
provides a mean metallicity for the cluster of $[Fe/H]=-1.55$. This metallicity
corresponds to $M_V(RR)=0.61\pm0.07$ if we consider the most recent (Hipparcos
based) calibration of the $M_V(RR)~vs.~[Fe/H]$ relation  \cite[see][and
references therein]{gisella}. Under this assumptions, the LS00 estimate of  
$<V^{RR}_0>$ corresponds to  $(m-M)_0=17.13\pm0.09$, in excellent agreement
with our result. 

The analysis of the effects of the adoption of our distance estimate on
dynamical models of the disruption of Sgr is clearly out of the scope of the
present paper. Just to provide an idea of the impact on some relevant scale we
note that (a) assuming the integrated apparent magnitude of Sgr estimated by
\citet[][hereafter M03, V$_0$=3.63]{steve} and our distance modulus, we obtain 
M$_V\simeq$-13.47, e.g. an increase in luminosity of 20\% with respect to the
results of M03 that assumed $(m-M)_0=16.90$; (b) the distances to stars in the
Sgr Stream estimated by M03 from the photometric parallax to M giants are
expanded by 10\% adopting our distance modulus.

\section*{Acknowledgments}

This research is partially supported by the italian {Ministero 
dell'Universit\`a e della Ricerca Scientifica} (MURST) through the COFIN grant
p.  2002028935, assigned to the project {\em Distance and stellar populations
in the galaxies of the Local Group}. Part of the data analysis has been
performed using software developed by P. Montegriffo at the INAF - Osservatorio
Astronomico di Bologna. We are grateful to P.~Bonifacio for useful comments
and  discussion. We also thank an anonymous referee for useful comments which
significantly improved our paper.
This work is based on observations made with the European
Southern  Observatory telescopes, using the Wide Field Imager, as part of the 
observing program 65.L-0463. Also based on data obtained from the  ESO/ST-ECF
Science Archive Facility. This publication makes also use of data products from
the Two Micron All  Sky Survey, which is a joint project of the University of
Massachusetts and  the Infrared Processing and Analysis Center/California
Institute of  Technology, funded by the National Aeronautics and Space
Administration and  the National Science Foundation. The support of ASI is also
acknowledged. This research has made use of NASA's Astrophysics Data System
Abstract Service.


\label{lastpage}

\end{document}